\documentclass[fleqn,10pt]{wlscirep}
\usepackage[utf8]{inputenc}
\usepackage[T1]{fontenc}
\usepackage{amsmath}
\usepackage{siunitx}
\usepackage{graphicx}
\usepackage{chemformula}
\graphicspath{ {./images/} }
\title{Compensating Atmospheric Channel Dispersion for Terahertz Wireless Communication}

\author[1]{Karl Strecker}
\author[1]{Sabit Ekin}
\author[1,*]{John F. O'Hara}
\affil[1]{School of Electrical and Computer Engineering, Stillwater, Oklahoma, 74078, USA}

\affil[*]{oharaj@okstate.edu}

\begin{abstract}
We report and demonstrate for the first time a method to compensate atmospheric group velocity dispersion of terahertz pulses. In ultra-wideband or impulse radio terahertz wireless communication, the atmosphere reshapes terahertz pulses via group velocity dispersion, a result of the frequency-dependent refractivity of air. Without correction, this can significantly degrade the achievable data transmission rate. We present a method for compensating the atmospheric dispersion of terahertz pulses using a cohort of stratified media reflectors. Using this method, we compensated group velocity dispersion in the 0.2-0.3~THz channel under common atmospheric conditions. Based on analytic and numerical simulations, the method can exhibit an in-band power efficiency of greater than 98\% and dispersion compensation up to 99\% of ideal. Simulations were validated by experimental measurements.
\end{abstract}
\begin{document}
\flushbottom
\maketitle
\thispagestyle{empty}

\section*{Introduction}
The capacity of wireless data links has risen exponentially in the last decade \cite{wells_2009, cherry_2004, song_nagatsuma_2011}, yet demand for higher data rates continues to increase. In the coming years, this increase is expected to accelerate \cite{ma_shrestha_moeller_mittleman_2018, cisco_2019}, even beyond the capacity of fifth-generation (5G) wireless \cite{dastjerdi_buyya_2016, cortes_bonnaire_marin_sens_2015, mumtaz_jornet_aulin_gerstacker_dong_ai_2017}. It has become apparent that the next generation of wireless communication technologies must push operating frequencies into the THz range to satisfy this predicted demand \cite{song_nagatsuma_2011, mumtaz_jornet_aulin_gerstacker_dong_ai_2017, chen_2019}. The development of any technology improving the bitrate-distance product or signal-to-noise ratio (SNR) of THz wireless links is an important step forward.

For THz wireless communications, there are two fundamental factors that limit achievable data rate and SNR. The most familiar is absorption, which limits SNR and is caused predominantly by rotational resonances of water vapor in the atmosphere \cite{Yang_2011, Harde_1997, Yang_2014, xin_2006}. These resonances also directly impact data rate by introducing group velocity dispersion (GVD). GVD arises when the many frequency components in a broadband signal propagate at different velocities due to the frequency-dependent refractive index of the channel medium. In the THz regime (0.1-10~THz), the atmosphere itself is one such dispersive channel \cite{hill_1988, Mandehgar_2014}. Because of GVD, single data bits, notionally represented by transform-limited THz pulses, spread temporally out of their assigned bit slot, superposing with bits in neighboring slots, resulting in inter-symbol interference (ISI). One solution to mitigate ISI is dilation of the bit slot, but this decreases the overall data rate \cite{agrawal_2011} and is clearly undesirable. 

GVD is an emerging problem in THz wireless systems. Unlike in microwave channels, water vapor resonances in the THz regime are very strong, and the available bandwidths are massive (60-180 GHz \cite{Mandehgar_2014}). Accordingly, it has been shown that GVD has a significant impact on THz wireless data rates \cite{Mandehgar_2014}. From previous work and our own calculations, reductions in data rate by a factor of 3 times or more are possible in single links over realistic signal ranges (1-20~km \cite{Wu_2017, Hirata_2010, Wang_2012}). Specifically, GVD will become limiting whenever the combination of link distance, atmospheric water vapor density, and channel bandwidth exceeds a certain minimum threshold, still to be defined. The majority of experimental THz links reported to date have not reached this threshold, either being long-range and low bandwidth \cite{Wu_2017, Hirata_2010, Wang_2012}, or short-range and high bandwidth \cite{Ducournau_2014, Antes_2012}. However, there is no fundamental limitation that prevents future technologies from exceeding this limit. When they do, GVD management will become a critical factor in overall channel performance, just as it was in earlier generations of fiber-optic systems \cite{agrawal_2011}.

One particular application where dispersion control may rapidly become critical is in the point-to-point distribution of ultra-high definition (UHD) and 3-D video \cite{Ducournau_2015}. For UHD video, uncompressed (real-time) data rates of up to 25~Gbps (giga bits per second) are required. For 3-D video, the data rate can reach 100~Gbps \cite{Ducournau_2014}. In a configuration analogous to fiber-optic systems with doped-fiber amplifiers \cite{agrawal_2011}, a series of point-to-point, wireless THz links might carry this high-bandwidth signal over tens of kilometers using repeater stations (which are generally more economical than full receivers). For such systems, dispersion would accumulate over the full distance of the link, easily causing a reduction in data rate by an order of magnitude or more. In this case, dispersion management becomes critical.

In this report, we present a method and device for compensating GVD in THz wireless communication channels that is highly effective, offers zero latency, consumes no power, and has low insertion loss. Using our approach in the 0.2-0.3~THz frequency channel, we demonstrate a compensation of up to 99\% of the GVD experienced by a transform-limited THz pulse propagating through 4~km of atmosphere with a water vapor density of $\rho_{\rm wv} = 10.37$~g/m$^3$ (60\% relative humidity at \SI{20}{\degreeCelsius}). Moreover, our approach can be implemented in a monolithic device that can simply be inserted into the THz beam to correct the signal GVD, and that has an insertion loss as low as 0.07~dB (over 98\% power efficient). To our knowledge, this is the first attempt to correct atmospheric GVD in the THz regime.

\section*{Results}
To compensate atmospheric dispersion, a device must be introduced into the THz beam that has GVD equal but opposite to that of the atmosphere. The resulting overall channel, comprised of both the atmosphere and the device, thus exhibits a cumulative GVD approaching zero. Our method uses a minimal structure of stratified dielectrics, arranged strategically in front of a high-reflector to form a multiple Gires-Tournois interferometer (MGTI) at THz frequencies. The dielectrics alternate between low and high refractive index, and the resonant trapping of frequencies within the structure introduces a frequency-dependent group delay. The optical thickness of each layer is tuned such that the overall GVD of the device matches the atmospheric opposite over a broad bandwidth. The THz wave experiences this opposite GVD during reflection from the MGTI.

\begin{figure}[ht]
\centering
\includegraphics[ width=0.75\textwidth]{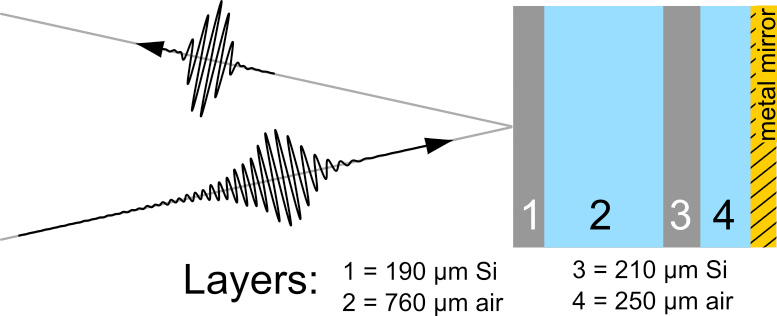}
\caption{Silicon-based MTGI design and concept. Temporally dispersed terahertz waves reflect from the MGTI and experience dispersion opposite that of the atmospheric channel, thus restoring their short pulsewidth.}
\label{fig:structure}
\end{figure}

\begin{figure}[ht]
\centering
\includegraphics[ width=0.75\textwidth]{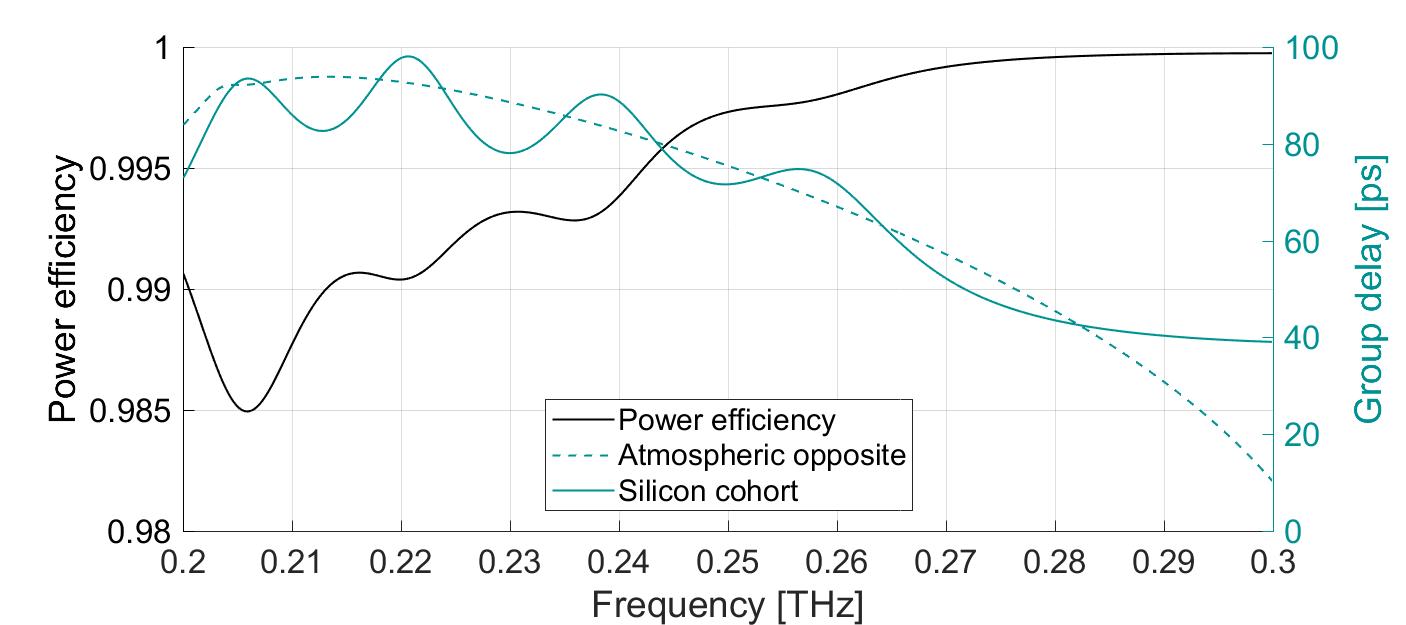}
\caption{Analytic performance of Si-MGTI cohort. Dashed line indicates the target group delay, the opposite of atmospheric group delay at $\rho_{wv} = 10.37$~g/m$^3$ for a 4~km propagation range. The black curve indicates the overall power efficiency. Plots show overall effects of one reflection from each MTGI of the cohort.}
\label{fig:simatch}
\end{figure}

The general MGTI structure is illustrated in Fig.\ref{fig:structure}. The parallel MGTI dielectric multilayers are readily modeled by means of characteristic matrices \cite{born_wolf_bhatia_2016} and the overall structures may be designed and rapidly optimized using genetic algorithms. For purposes of design and simulation, it is convenient to use group delay instead of GVD, since the former is specific to a set of channel conditions, whereas GVD is normalized to channel length. The curve labeled “Atmospheric Opposite” in Fig. \ref{fig:simatch} shows the ideal group delay profile for compensating a 4~km path length through an atmosphere with uniform water vapor density $\rho_{\rm wv} = 10.37$~g/m$^3$ (60\% relative humidity at \SI{20}{\degreeCelsius}) over the entire 100~GHz band centered at 0.25~THz. As detailed in the methods section, the absorption and refractivity of the atmosphere were calculated using Molecular Response Theory \cite{Mandehgar_2013, Harde_1997, Yang_2014} and water vapor resonance line data from the HITRAN database \cite{gordon_2017}.

While Fig.~\ref{fig:structure} shows only one design, two different MGTIs were actually employed together to compensate the above described dispersion. Using a cohort of two MGTIs increases the effectiveness of compensation by improving the fit to the atmospheric opposite over a larger bandwidth. Each MGTI was designed with alternating layers of high-resistivity silicon (Si) and air. The layer thicknesses were obtained by using a genetic algorithm to compute the optimal characteristic matrix for a cohort of two MGTIs in reflection. The optimal layering for the first MGTI was ${190\;\mu}$m of silicon, ${760\;\mu}$m of air, ${210\;\mu}$m of silicon, ${250\;\mu}$m of air, and finally the reflector. Thickness variations of ${\pm 5\; \mu}$m had minor effects to overall performance. The silicon was assumed to be lossless with a refractive index of ${n_{\rm Si}=3.418}$. For the second MGTI, the layering was ${210\;\mu}$m of silicon, ${800\;\mu}$m of air, ${260\;\mu}$m of silicon, ${90\mu}$m of air, reflector. In operation, the THz wave would reflect once from each MGTI to achieve full compensation. The cohort reflects more than 98\% of incident power (0.03~dB insertion loss per MGTI per reflection) where conductor losses in the high-reflector constituted the majority loss mechanism. The cohort design also resulted in a good fit to the target group delay curve, as can be seen from Fig.~\ref{fig:simatch}.

\begin{figure}[ht]
\centering
\includegraphics[ width=0.75\textwidth]{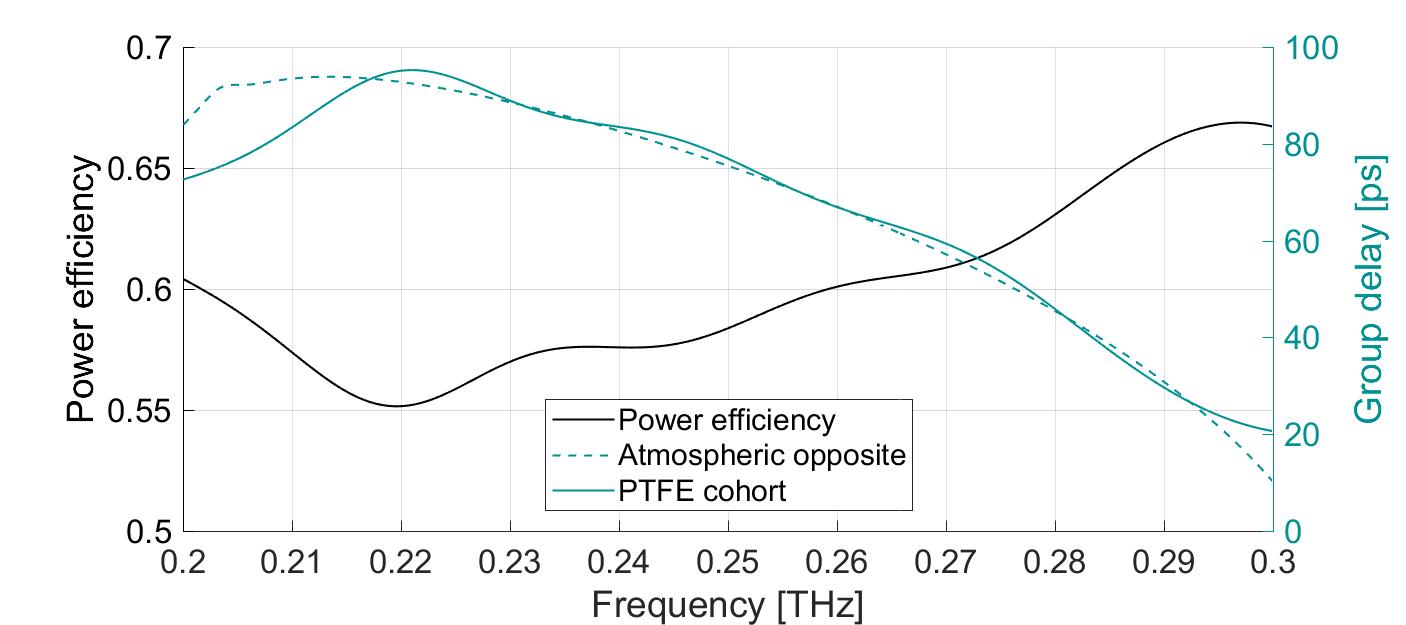}
\caption{Analytic performance of PTFE-MGTI cohort. Dashed line indicates the target group delay, the opposite of atmospheric group delay at $\rho_{\rm wv} = 10.37$~g$/m^3$ for a 4~km propagation range. The black curve indicates the overall power efficiency. Plots show overall effects of four reflections from each MTGI of the cohort.}
\label{fig:ptfematch}
\end{figure}

As a second demonstration, a cohort of MGTIs based on polytetrafluoroethylene (PTFE) layers was also designed. The first MGTI of the cohort had a layering of ${480\;\mu}$m of PTFE, ${380\;\mu}$m of air, ${510\;\mu}$m of PTFE, ${250\;\mu}$m of air, and finally the reflector. The second MGTI of the cohort had a layering of ${430\;\mu}$m of PTFE, ${750\;\mu}$m of air, ${540\;\mu}$m of PTFE, ${290\;\mu}$m of air, then the reflector. As before, the cohort was designed to compensate the dispersion produced by 4~km of propagation through the atmosphere with $\rho_{wv} = 10.37$~g/m$^3$ (60\% relative humidity at \SI{20}{\degreeCelsius}). Note that PTFE has a lower index than silicon (a constant index ${n_{\rm PTFE}=1.42+i0.0032}$ was assumed based on measurements), which reduces the reflections at the layer boundaries, resulting in less overall dispersion. This allowed us to more precisely match the atmosphere’s dispersion profile for varying water vapor densities at the expense of cumulative dispersive power.

PTFE absorptive losses are not large in the THz range, but they accumulate when waves traverse the layers of the MGTIs multiple times, leading to greater overall insertion loss than a silicon-based device. Because of this, both material and reflector losses must be accounted for. Despite this added loss, the PTFE cohort is still highly effective at compensating atmospheric dispersion over shorter distances. This cohort of two MGTIs reflects more than 86\% of incident power for a single pass through the cohort (0.3~dB insertion loss per MGTI per reflection). However, since its dispersive power is about 25\% of the Si-based cohort, the THz beam must reflect from each PTFE-MGTI four times to compensate the target 4~km channel, which lowers the overall cohort power efficiency to 55\% (minimum) or 2.6~dB insertion loss total, as shown in Fig.~\ref{fig:ptfematch}. An advantage of the PTFE-cohort, apparent from Fig.~\ref{fig:ptfematch}, is that it exhibits a better fit to the target group delay curve.

\begin{figure}[ht]
\centering
\includegraphics[ width=0.85\textwidth]{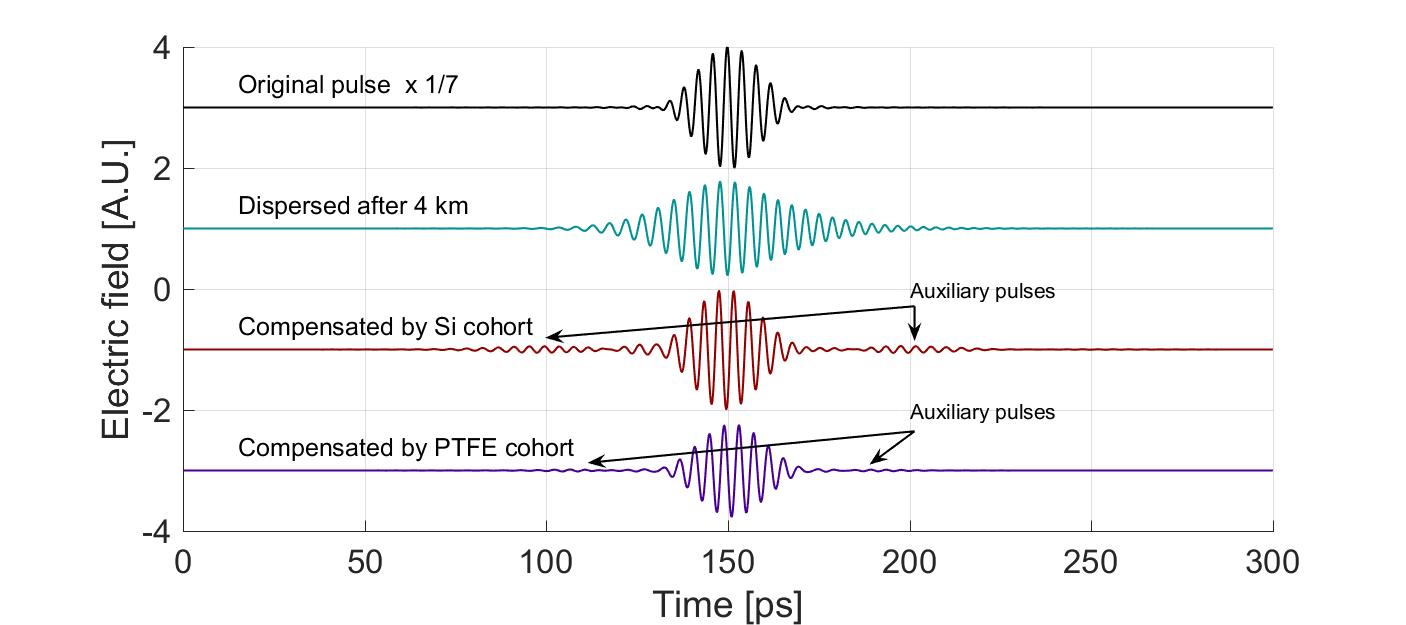}
\caption{Time-domain terahertz pulses before and after propagation through atmosphere and compensators. (Black) Original pulse, (red) pulse after propagation through atmosphere, (blue) pulse after Si-MGTI cohort compensation, (green) pulse after PTFE-MGTI cohort. All waveforms are shown to scale except the original pulse, which is scaled down in amplitude by 7 times to aid display. Auxiliary pulses are evident in compensated waveforms, as shown.}
\label{fig:compensated}
\end{figure}

\begin{figure}[ht]
\centering
\includegraphics[ width=0.5\textwidth]{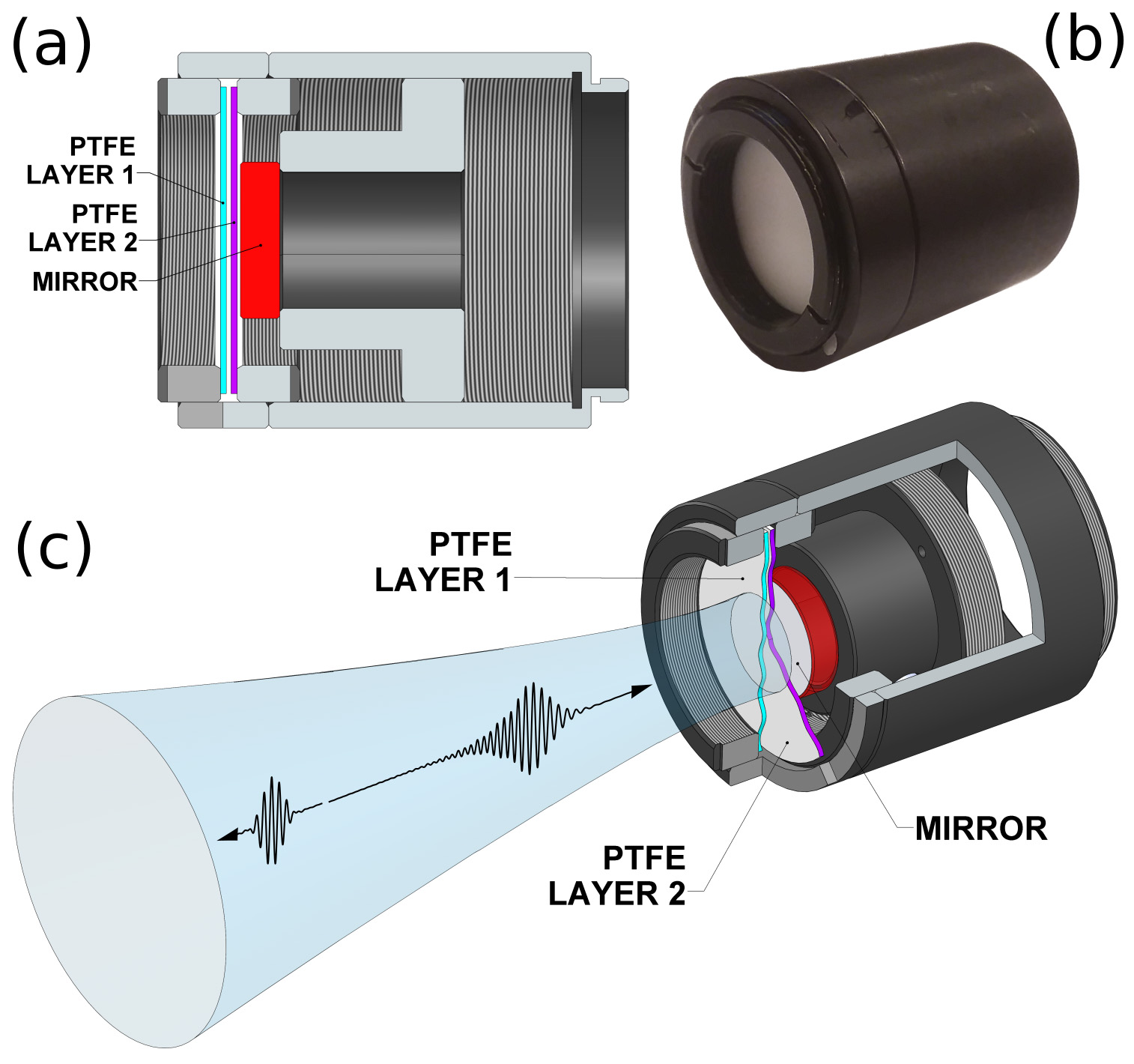}
\caption{A prototype terahertz dispersion compensating MGTI. (a) cross-section drawing of designed and fabricated MGTI, (b) fabricated MGTI, with PTFE first layer visible (c) scale cutaway drawing of fabricated MGTI showing internal layering. Blue tube depicts focused input/output terahertz beam.}
\label{fig:sample}
\end{figure}

To further demonstrate the MGTI cohorts, we calculated the time-domain waveforms for THz pulses before and after compensation. Again, we assumed plane-wave propagation of a THz pulse through the atmosphere at $\rho_{\rm wv} = 10.37$~g/m$^3$ over a distance of 4~km. The initial transform-limited pulse was given a raised cosine spectrum centered at 0.25~THz, with a full-width-half-max (FWHM) bandwidth of 0.05~THz. After propagating through the atmosphere, this pulse was dispersed to 175\% of its original width by GVD and reduced in amplitude by 9.1 times due to water vapor absorption losses, as shown by the electric-field waveforms in Fig.~\ref{fig:compensated}. For determination of the pulse width, we assumed an approximately Gaussian profile of the pulse both before and after dispersion. 

This dispersed pulse was compensated by both the Si- and PTFE-based MGTI cohorts. As before, compensation was achieved by either one (Si) or four (PTFE), normal-incidence reflections off the cohort. These atmospheric conditions represent a natural transition point between the silicon and PTFE approaches. The silicon cohort cannot be used for less compensation because there can be no less than one reflection, while the PTFE cohort is less desirable for more compensation because of accumulating losses. The analytic transfer functions of the cohorts were calculated and multiplied by the dispersed pulse spectrum in the frequency domain, then inverse Fourier transformed to obtain the time-domain output waveforms. 

As shown in Fig.~\ref{fig:compensated}, the pulse went from 175\% to 105\% (for both Si and PTFE) of its original width. This significant reduction in pulse width corresponds to a 66\% increase in spectral efficiency, ${\eta_{eff}}=B/{\Delta}{f_{\rm ch}}$ \cite{agrawal_2011}, where ${\Delta}{f_{\rm ch}}$ is the pulse bandwidth, and $B$ is the bit rate, which is inversely proportional to the pulse width. These numbers do not fully describe the effectiveness of the approach. Even a perfect dispersion compensation scheme cannot fully compensate for atmospheric dispersion, because some signal bandwidth is irreversibly lost due to absorption \cite{Mandehgar_2015}. When this is accounted for, the Si-MGTI and PTFE-MGTI cohorts both achieve >99\% of the theoretically possible dispersion compensation, but the PTFE cohort is better, nearly reaching perfect compensation (99.8\% by calculation).

It is worth noting that the missing $\sim1$\% of possible compensation for the Si cohort is manifested in the presence of ``auxiliary pulses'' on the compensated waveforms of Fig.~\ref{fig:compensated}. These pulses contain very little power and are not large enough to significantly contribute to ISI. Their presence may be linked back to the frequency-dependent oscillations in the cohort’s group delay around the target group delay (See Fig \ref{fig:simatch}). In other words, they result from the non-zero derivative of the sum of group delays of the atmosphere and the MGTIs. The auxiliary pulses can be minimized by using a lower index material (such as PTFE) to construct the MGTIs. This reduces the lifetime of resonances within the MGTI layers, thus reducing the amplitude of the group delay oscillations. For example, both the group delay oscillations (Figs.~\ref{fig:simatch} and \ref{fig:ptfematch}) and the auxiliary pulses (Fig.~\ref{fig:compensated}) are much smaller in the PTFE cohort than in the Si cohort. Auxiliary pulses may also be suppressed by the addition of more layers to the MGTIs, or by the addition of more MGTIs to the cohort. However, given the exceptionally good match achieved with a small number of MGTIs and dielectric layers, these adjustments appear unnecessary. For longer range systems, such strategies might be employed with Si-based MGTIs to obtain every advantage in both optimized dispersion compensation and low loss.

\section*{Experimental validation}
The analytic calculations were supported by both full wave simulations and experimental measurements, the latter detailed here. It was not possible to experimentally implement a 4~km channel with uniform and tuned atmospheric properties. But the absorptive and dispersive behaviors of the atmosphere are already well-understood in great detail and the models have been verified by experimentation \cite{Yang_2014, OHara_2018}. Therefore, to validate the concept, it is sufficient to demonstrate only that an MGTI-based compensator can produce the target “atmospheric opposite” group delay in the bandwidth of interest. We elected to fabricate and measure one MGTI of the PTFE-cohort described above and measure it with reflection-mode terahertz time-domain spectroscopy (THz-TDS). This MGTI has an ideal structure of ${480\;\mu}$m of PTFE, ${380\;\mu}$m of air, ${510\;\mu}$m of PTFE, ${250\;\mu}$m of air, and finally an aluminum mirror. Figure~\ref{fig:sample}(b) shows the actual fabricated device alongside a schematic of its inner structure in Fig. \ref{fig:sample}(c). The MGTI diameter was made large enough ($\sim$12~mm) to avoid clipping the THz beam focused onto its surface. Further fabrication and measurement details are given in the methods section below.

\begin{figure}[ht]
\centering
\includegraphics[ width=.8\textwidth]{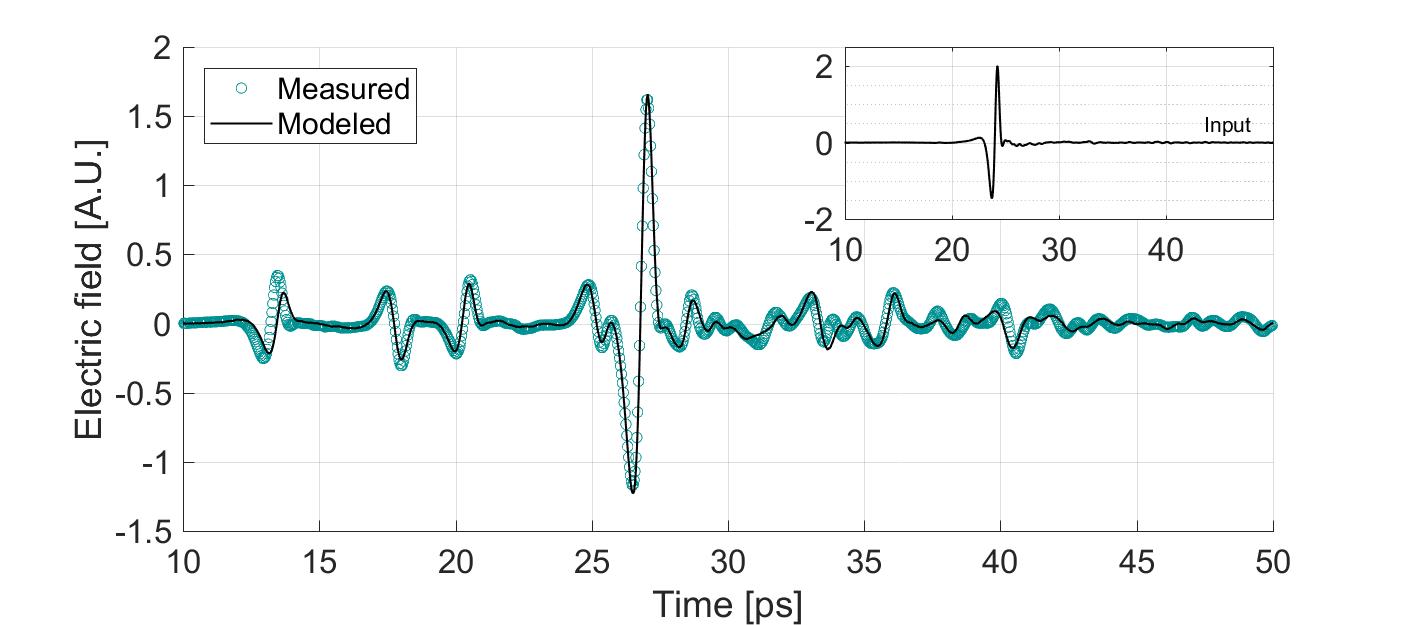}
\caption{Measured and predicted time-domain waveforms for THz pulses reflected from the PTFE MGTI shown in Fig.~\ref{fig:sample}. The inset shows the measured reference pulse, obtained by removing the two PTFE layers in front of the reflector.}
\label{fig:timedomain}
\end{figure}

The measured and predicted time-domain traces are shown in Fig. \ref{fig:timedomain}. This match is presented in the time-domain to better illustrate that the waveforms are nearly identical. This proves that the MGTI is behaving almost exactly like the model over the \emph{entire} measured frequency range (~0.1-2.5~THz), both in phase and amplitude. A notable discrepancy between the measured and predicted waveforms is found at the earliest (leftmost) feature on the plot. This feature is the reflection off the first PTFE layer of the MGTI, and its slight time misalignment suggests that this layer was fabricated thinner than intended.

\begin{figure}[ht]
\centering
\includegraphics[ width=0.75\textwidth]{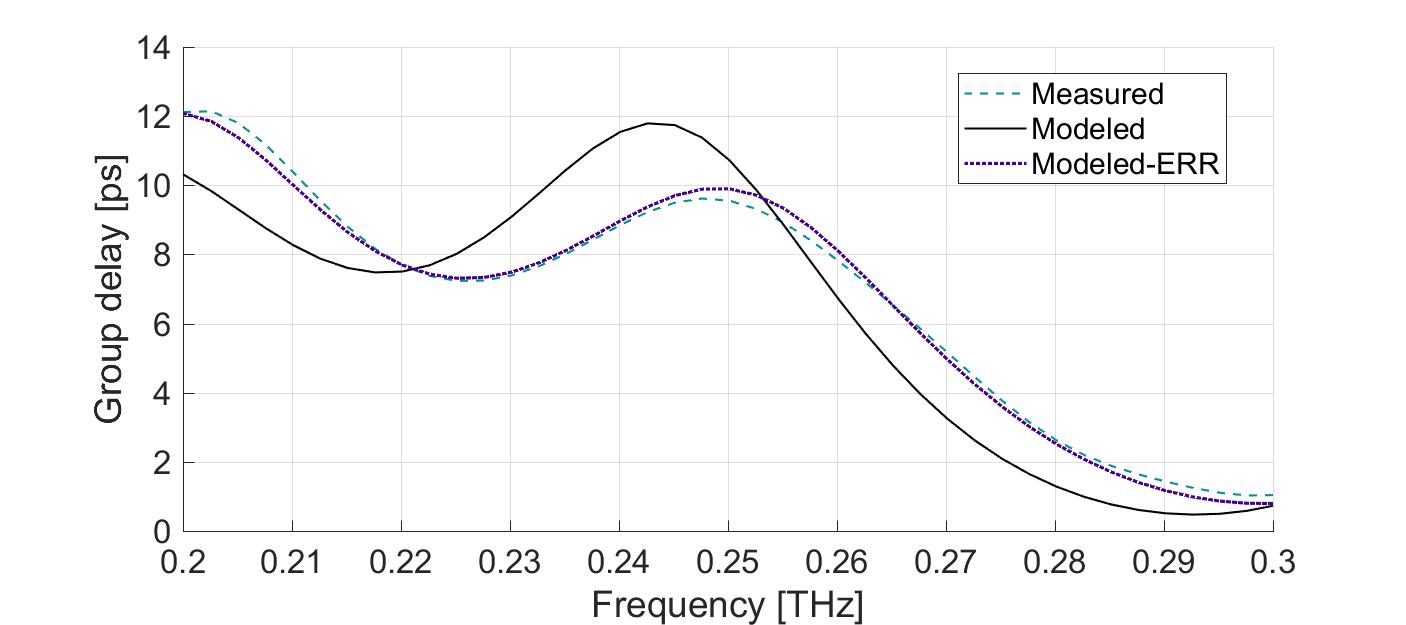}
\caption{Measured and predicted group delay for THz waves reflected from the PTFE MGTI shown in Fig.~\ref{fig:sample}}
\label{fig:freqdomain}
\end{figure}

The measured data can be analyzed in the frequency domain to extract the group delay and compare to model predictions, as shown in Fig.~\ref{fig:freqdomain}. Again, there is good agreement between the modeled and measured values, which confirms both our analytical calculations and the validity of the GVD compensation approach. We note that these curves represent the effect of only one reflection from one MGTI of the cohort, hence the graphs of Figs. \ref{fig:ptfematch} and \ref{fig:freqdomain} are different. Discrepancies between measured and predicted results arise primarily because our fabrication techniques could not exactly produce the desired layer thicknesses nor uniformity. Specifically, the first PTFE layer appears to have been fabricated as ${450\;\mu}$m, or ${30\;\mu}$m thinner than intended. When the model is modified to account for this error, the predicted and measured group delay profiles are in excellent agreement, as shown by the trace labeled “Modeled-ERR” in Fig.~\ref{fig:freqdomain}. The experimental results confirm that our analytic calculations are accurate, and that such devices are readily designed and fabricated.

\section*{Discussion}
Since the THz wireless channel would generally be dynamic, both in terms of atmospheric properties (weather) and signal range, it is important to address the adaptability of our approach to changing channel conditions. The dispersion of an MGTI cohort can be dynamically adapted most easily by altering the number of times the THz beam is reflected from it. By mechanically adjusting the angle and position of the individual cohort members, the dispersed signal can be made to undergo more or fewer reflections from each MGTI as desired, which corresponds to a discrete increase or decrease in the level of dispersion achieved. By using a low-index dielectric MGTI in the cohort, the resolution of these discrete changes can be very fine, making it possible to compensate effectively a continuous range of changing atmospheric conditions and ranges. Even though changing the number of reflections would be a slow tuning procedure, it is more than sufficient for this application because channel conditions also change quite slowly.

\begin{figure}[ht]
\centering
\includegraphics[ width=0.75\textwidth]{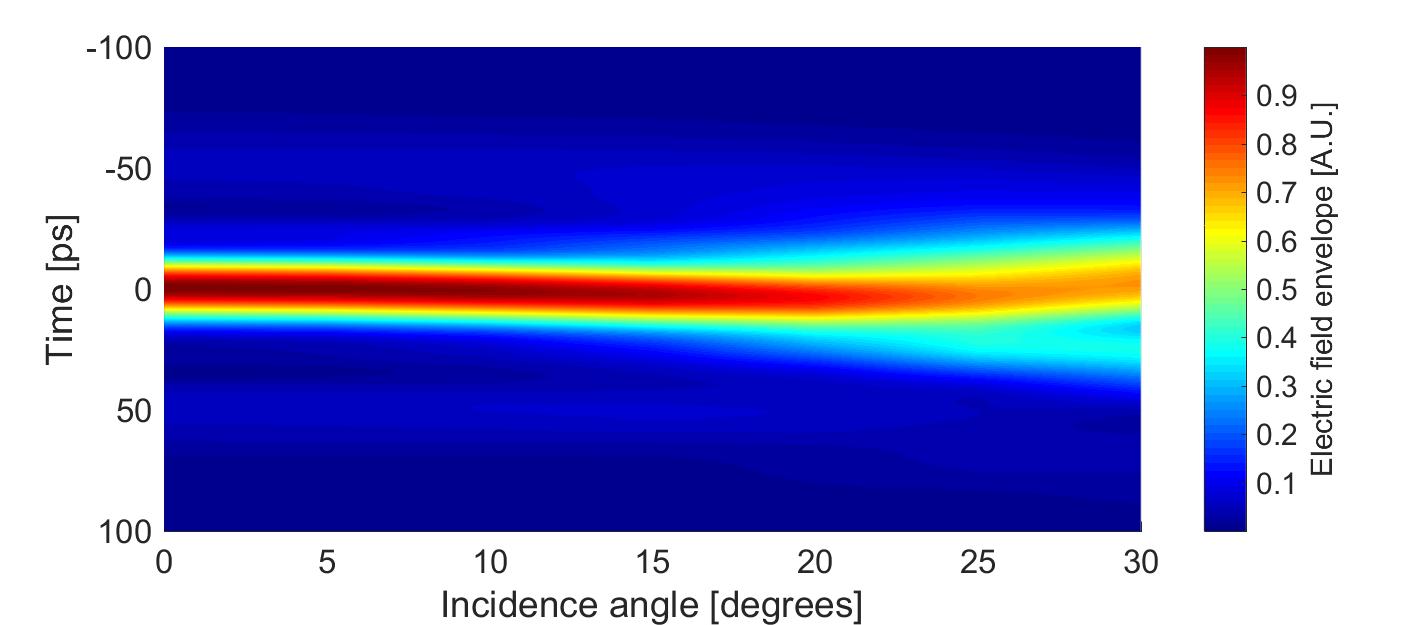}
\caption{Dependence of compensated pulse width on obliquity angle for Si-MTGI cohort, assuming p-polarization. The pulse begins to noticeably disperse in time when the obliquity angle exceeds ${15^{\circ}}$.}
\label{fig:offangle}
\end{figure}

Altering the physical geometry of the signal path to either add or reduce reflections will also mean changing the angle at which the wave is incident on each MGTI. However, the complex reflection coefficient of a stratified dielectric is a function of incidence angle and polarization\cite{born_wolf_bhatia_2016}. Thus, there is a limit to how obliquely the signal may strike each MGTI before the cohort will no longer adequately compensate the atmospheric dispersion. Our calculations show that the silicon cohort described above will still achieve 90\% or greater of the possible dispersion compensation as long as the incidence angle for all MGTIs is less than ${15^{\circ}}$ off normal in either direction, for both s and p polarizations. This is sufficiently flexible for the mechanical tuning techniques described above, and it also illustrates the excellent robustness of the technique to practical alignment errors. The dependence of compensated pulse shape on incidence angle for the Si cohort is shown in Fig. \ref{fig:offangle} for a p-polarized (worst case) incident pulse.

In conclusion, we have presented an effective method for compensating atmospheric dispersion in long-distance THz wireless links to thereby maximize data transmission rates to their fundamental limit. To our knowledge, this device is the first of its kind, and this work represents an important step toward the implementation of point-to-point and point-to-mobile THz wireless links. Furthermore, compensating devices such as the ones presented here are highly effective in terms of both complete dispersion management and low loss, inexpensive to fabricate, and are surprisingly tolerant to manufacturing errors. Though not discussed in this paper, they may also be easily optimized for oblique incidence, which allows for flexibility in how such devices are physically incorporated into THz communication systems.

\section*{Methods}
\subsection*{Atmospheric modeling}
The atmosphere was modeled using molecular response theory (MRT), in conjunction with molecular resonance data from the HITRAN database. For each \ch{H2O} molecular resonance in the database, the frequency, resonance strength, and broadening factor was extracted. Resonances too weak or too far distant to influence the frequency range of interest were ignored. MRT was used to model the broadened lineshapes of the remaining resonances. MRT, rather than van-Vleck-Weisskopf or Full Lorentzian models, is used to model the broadened resonance lineshapes because MRT produces a superior match to experimental data \cite{OHara_2018, Yang_2014, Harde_1997}. The exact shape of the broadened resonance lines produced by the MRT model depends heavily on the temperature, pressure, and water vapor density of the atmosphere, so these parameters must be specified before modeling. After the broadened lineshapes are determined, they are summed up over the frequency range of interest to produce the frequency-dependent transfer function of the atmosphere. Finally, multiplying the transfer function of the atmosphere by the Fourier transform of the input waveform (which is assumed to be a plane wave) yields the frequency-domain representation of the dispersed waveform. Converting this back to the time domain gives the dispersed pulse.

\subsection*{Experimental setup}
\begin{figure}[ht]
\centering
\includegraphics[ width=0.5\textwidth]{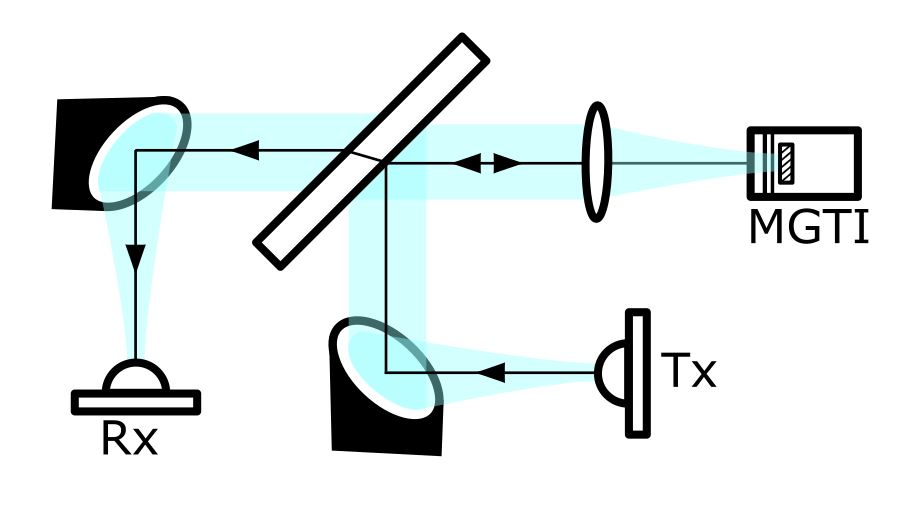}
\caption{Reflection-mode THz-TDS setup for normal-incidence, phase-coherent measurements. The system is confocal and produces a frequency-independent beam waist with a planar phase front at the MGTI sample. The blue shaded areas illustrate the terahertz beam profile (not to scale). Labels ``Tx'' and ``Rx'' refer to the terahertz transmitter and receiver.}
\label{fig:schematic}
\end{figure}

Our reflection-mode THz-TDS setup permits phase-coherent measurements of the sample and reference at normal incidence and is illustrated in Fig.~\ref{fig:schematic}. It is modified from the standard transmission setup described in the reference \cite{VanExter_1990}. Generated terahertz pulses are collimated by an off-axis paraboloidal mirror to a 3~mm thick high-resistivity float-zone silicon beamsplitter. From here, they are reflected toward and focused onto the sample at normal incidence using a polyethylene lens. At the sample, the frequency-independent beam waist diameter is $\sim 7$~mm. After reflecting off the sample, the broadband beam is again collimated by the polyethylene lens and then passes through the beamsplitter. It is finally focused into the THz receiver by another off-axis paraboloidal mirror. The entire system is confocal to maximize power transfer and ensure a frequency-independent beam waist at the sample. 

The thickness of the silicon beamsplitter is chosen to be large so that the initial pulse and subsequent features in the time-domain are easily separable from later echo pulses, which are caused by multiple reflections within the beamsplitter. If the beamsplitter is too thin, the time-domain features introduced by the sample will overlap with the features introduced by the beamsplitter, significantly increasing the difficulty of extracting the sample parameters. Obtaining the transfer function of the MGTI requires knowledge of the spectrum of the incident wave. The incident wave is measured by removing the two PTFE layers in front and measuring only the aluminum mirror of the MGTI sample. The MGTI is mounted such that the mirror does not move when the PTFE layers are added between sample and reference measurements, thus establishing a fixed phase reference for proper determination of the transfer function.

\subsection*{Fabrication of the sample}
The sample was constructed from 20~mm diameter PTFE sheets, manually spaced using the apparatus shown in Fig.~\ref{fig:sample}. PTFE sheets of the desired thicknesses were not commercially available, and thus had to be fabricated by hand. Consequently, though the fabricated layer thicknesses and spacings were close to the intended values, the thickness of the layers and spacings were not ideal. All layers were within the $\pm 5\;\mu$m tolerance, with the exception of the first PTFE layer, which measurements indicated was about 30 microns thinner than intended. According to our numerical analyses, these fabrication errors – with the exception of the first layer – are small enough to not significantly affect the group delay profile of the sample.

\section*{Data Availability}
The datasets generated and analysed during the current study are available from the corresponding author on reasonable request.

\bibliography{references}

\section*{Acknowledgements}

The authors gratefully acknowledge Prof. Daniel Grischkowsky for his helpful perspectives and suggestions on this manuscript.

\section*{Author contributions}
J.O. and K.S. conceived and/or performed the dispersion compensation approach, the theoretical analysis and simulations, the sample design methods, and the measurement method.  K.S. fabricated the sample and conducted the measurements.  K.S., J.O., and S.E. analysed the data and wrote the manuscript.

\section*{Competing interests}


The authors declare no competing interests.

\end{document}